\documentclass[12pt,a4paper]{article}

\usepackage{amsmath}

\def\hybrid{\topmargin -20pt    \oddsidemargin 0pt
        \headheight .4in \headsep 0pt
        \textwidth 6.25in       
        \textheight 9.1in       
        \marginparwidth .875in
        \parskip 5pt plus 1pt   \jot = 1.5ex}

\hybrid

\numberwithin{equation}{section}
\numberwithin{table}{section}

\newcommand{\ap}{{\alpha'}}
\newcommand{\f}{\mathbf{27}}
\newcommand{\af}{\overline{\mathbf{27}}}
\newcommand{\nn}{\nonumber}

\begin{document}

\begin{titlepage}

  \begin{center}
    {\Large\textbf{Moduli Stabilisation in Heterotic Models
        \\[.2cm] with Standard Embedding}}

    \vskip1cm

    \textbf{Andrei Micu}
    \vskip.5cm
    \textit{Department of Theoretical Physics \\ National Institute
      for Physics and Nuclear Engineering -- Horia Hulubei \\
      Str. Atomi\c{s}tilor 407, P.O.~Box MG-6, M\u{a}gurele, 077125 Romania}\\
    \texttt{amicu@theory.nipne.ro} 

    \vspace{2cm}
    \textbf{Abstract}
  \end{center}

  \noindent In this note we analyse the issue of moduli stabilisation
  in 4d models obtained from heterotic string compactifications on
  manifolds with $SU(3)$ structure with standard embedding. In order
  to deal with tractable models we first integrate out the massive
  fields. We argue that one can not only integrate out the moduli
  fields, but along the way one has to truncate also the corresponding
  matter fields. We show that the effective models obtained in this
  way do not have satisfactory solutions. We also look for stabilised
  vacua which take into account the presence of the matter fields. We
  argue that this also fails due to a no-go theorem for Minkowski
  vacua in the moduli sector which we prove in the end. The main
  ingredient for this no-go theorem is the constraint on the fluxes
  which comes from the Bianchi identity.

\vfill

\noindent {November 2009}
  
\end{titlepage}

\section{Introduction}

In recent years there have been considerable progress in the field of
moduli stabilisation in string compactifications especially due to the
idea of flux compactification (for reviews see
\cite{Grana,DK}. However, moduli stabilisation in realistic
models is far from being well understood. This is an important issue
in trying to construct viable models from string theory as,
stabilising moduli by using fluxes may influence the matter sector in
a significant way. It is not really clear in general whether the
moduli stabilisation sector is independent of the matter sector and if
this is not the case, what is the relation between these sectors. The
best one can achieve so far is to analyse this kind of questions on a
case by case basis.

Moduli stabilisation in realistic models\footnote{By realistic here we
  mean models which include a matter sector.} was studied mostly in
type II or F-theory models, mainly due to the better control on the
fluxes in these theories \cite{BCKMQ,BMP1,BMP2,CMQ}.  As it is well
known, the lack of ordinary fluxes in heterotic string is a big
challenge for the moduli stabilisation problem. The presence of fluxes
inevitably deforms the compactification geometry away from Calabi--Yau
manifolds \cite{AS}--\cite{BT} and one has to deal with non-K\"ahler
manifolds. Moreover, the simple solutions with fluxes in type IIB are mapped
via duality to torsional backgrounds in heterotic theories
\cite{DRS,BS}.  Therefore, one has to understand the theories which
result from compactifications on backgrounds with torsion. Using a
supergravity approach, the effective theory which comes from heterotic
string compactification on certain manifolds with $SU(3)$ structure
was derived in \cite{GLM1,GLM2}. The background consists of half-flat
manifolds, which have been encountred in the context of mirror
symmetry \cite{GLMW,GM}, and generalisations thereof
\cite{AFTV,tomasiello,thomas,GLW}. In \cite{AM} we started to consider
the moduli stabilisation problem in the models of \cite{GLM2} and here
we shall continue along the same route in that we make a supergravity
analysis of the moduli stabilisation problem in models which contain
matter fields. The particular setup we shall consider is that of the
standard embedding. To be more precise, we have in mind as a starting
point, heterotic string compactifications on Calabi--Yau manifolds
with standard embedding. The actual background is obtained by
deforming the Calabi--Yau manifold into a manifold with $SU(3)$
structure as prescribed in \cite{GLMW} so that we still retain
cont-role on the compactification procedure and we are able to write
down an effective action in four dimensions.\footnote{For another view
  on the standard embedding in heterotic string compactifications on
  manifolds with $SU(3)$ structure see \cite{tibra}.}

The good part of this approach is that the calculation holds for any
Calabi--Yau manifold and its corresponding deformation to a manifold
with $SU(3)$ structure. This is mainly because of the standard
embedding we use for solving the Bianchi identity. In particular it
can be applied to the interesting models with three generations
recently found in \cite{CBD}.
On the other hand,
as we shall see later in the paper, the format of the theory is fairly
rigid and the connection between the matter and moduli sector (which is
again due to the standard embedding procedure) makes it difficult --
if not impossible -- to find good models with stabilised moduli. 

Finding solutions in supergravity coupled to a few chiral fields is
already a challenging task. For the case at hand we would have to deal
with large number of fields, both neutral and charged and therefore a
complete classification of the solutions is far outside our
reach. Instead, we shall find effective models -- which are
easier to solve -- by integrating out certain massive fields. It turns
out that the masses for the matter fields and for the moduli are of
the same order, and therefore, if we want to integrate our certain
moduli fields we should also integrate out the corresponding matter
fields. This will essentially leave us with very simple models which
have very small flexibility of finding good solutions. 

One important aspect which we will have in mind is that of the
stabilisation of the dilaton. The superpotentials we consider do not
depend on the dilaton and therefore the only possibility which is
left, is to consider its stabilisation via non-perturbative effects
such as gaugino condensate in the hidden sector. For a realistic
stabilisation we would need to obtain a small value for the
superpotential after stabilising all other fields. This is actually
the main challenge in our analysis as it was realised long ago
\cite{DIN2,DRSW}. Finding small values for the flux superpotential
directly seems difficult because of the fact that the effective models
we work with have a too simple superpotential which can not be made
small and at the same time ensure large values for the moduli fields
which are required by the consistency of the compactification. Another
possibility we consider is to find small superpotentials which are
generated from the matter fields. However this requires a vanishing
flux-superpotential which we show it is impossible to obtain in the
context of the standard embedding.

The organization of the paper is as follows. In section \ref{model} we
shall review the model under consideration. This was originally
obtained in \cite{GLM2}.  In section \ref{masses} we shall analyse the
issue of the masses of the moduli and matter fields. We shall make the
remark that the masses for matter and moduli fields are related and
therefore if we wish the moduli to get large masses, the same will
happen with certain mater fields. Therefore when integrating out
massive moduli we would have to integrate out at the same time the
massive matter fields as well. In section \ref{effective} we study two
types of effective models, the first one containing only K\"ahler
moduli and the second containing only complex structure moduli. We
shall argue that none of these models has satisfactory
solutions. Finally in section \ref{matterw} we shall analyse the
possibility of obtaining a small superpotential only using the matter
sector. For this we will need to find solutions with vanishing flux
superpotential, but as we shall claim in this section this turns out
to be impossible provided one satisfies the Bianchi identity. In
section \ref{conclusions} we present our conclusions.

\section{The general model}
\label{model}

As explained in the Introduction we shall stay in the realm of
heterotic string compactifications on manifolds with $SU(3)$ structure
with standard embedding. To be more precise, we consider Calabi--Yau
compactifications of the $E_8 \times E_8$ heterotic string with
standard embedding and deform the compactification manifold to one
with $SU(3)$ structure \cite{GLM2}. The effective four-dimensional
theory consists of $N=1$ supergravity coupled to a $E_6$ super
Yang-Mills theory\footnote{There is also a hidden sector consisting of
  the second $E_8$ factor of the ten-dimensional gauge group, but this
  will be irrelevant for most of the issues discussed here.} and to a
certain number of chiral (super)fields. The chiral fields -- which we
shall be concentrating on in the following -- come in two categories:
\begin{itemize}
\item neutral fields: the axio-dialton, $S$, the K\"ahler moduli, $T^i$
  and the complex structure moduli $Z^a$.
\item charged (matter) fields: transforming in the representation
  $\f$, $D^a$ and in $\af$, $C^i$, of $E_6$ (we have suppressed the $E_6$ index)
\end{itemize}

On top of these fields one also has to consider the so called  bundle
moduli which parameterise the possible deformations of the gauge
bundle. However, dealing with these fields is s lot more difficult and
we shall not consider them in the following. 

The K\"ahler moduli, as well as the matter fields in $\af$ are in one
to one correspondence with the $(1,1)$ forms on the compactification
manifold and therefore they carry an index $i = 1,\ldots, h^{1,1}$ while
the complex structure moduli and the matter fields in $\f$ are in one
to one correspondence with the $(2,1)$ forms and therefore carry an
index $a = 1, \ldots, h^{2,1}$, where $h^{1,1}$ and $h^{2,1}$ are dimensions
of the corresponding cohomology groups. The kinetic terms for the chiral
fields are given in terms of the K\"ahler potential 
\begin{equation}
  \label{Kgen}
  K(S,T,Z,C,D) = K_0 (S,T,Z) + \ap K_1(T,Z, C,D) \; ,
\end{equation}
where
\begin{eqnarray}
  \label{K0}
  K_0 & = & - \log{(S + \bar S)} - \log{\tfrac16[\mathcal{K}_{ijk}(T^i + \bar
      T^i) (T^j + \bar T^j) (T^k + \bar T^k)]} \\
    & & - \log{\tfrac16[\tilde{\mathcal{K}}_{abc}(Z^a + \bar 
      Z^a) (Z^b + \bar Z^b) (Z^c + \bar Z^c)]} \; \nonumber \\
    \label{K1}
    K_1 & = & 4 e^{(K_{cs} - K_K)/3} g_{ij} C^{i \bar P} \bar
    C^{j_{\bar P}} + e^{(K_K - K_{cs})/3} g_{a \bar b} D^{aP} \bar
      D^{\bar b}_P - 2 \left(K_i K_a C^i_P D^{aP} + c.c. \right) \; .
\end{eqnarray}
So far, this is the usual result of Calabi--Yau compactifications of
the heterotic string with standard embedding. The fact that we
consider a manifold with $SU(3)$ structure and background fluxes shows
up in the superpotential which reads
\begin{equation}
  \label{Wgen}
  W(T,Z,C,D) = W_0(T,Z) + \ap W_1(Z,C,D) \; , 
\end{equation}
where
\begin{eqnarray}
  \label{W0}
  W_0 & = & i \left( \xi + i e_i T^i \right) + \left( \epsilon_a +
    ip_{ia}T^i \right)Z^a + \tfrac{i}{2} \left( \mu^a + i q_i^a T^i \right)
    \tilde{\mathcal{K}}_{abc} Z^b Z^c \nonumber \\
    & & + \tfrac16 \left( \rho + i r_i T^i \right)
    \tilde{\mathcal{K}}_{abc} Z^a Z^bZ^c \; ,\\ 
    \label{W1}
    W_1 & = & 2 \left[i p_{ia} + (\tfrac{i}2 r_i Z^a - q_i^a)
      \tilde{\mathcal{K}}_{abc} Z^b\right] C^i D^c \nonumber \\
    & & - \tfrac{1}{3} \left[\mathcal{K}_{ijk} j_{\bar P \bar R \bar S}
      C^{i \bar P} C^{j \bar R} C^{k \bar S} + 
      \tilde{\mathcal{K}}_{abc} j_{PRS} D^{a P} D^{b R} D^{c S}
    \right] \; .
\end{eqnarray}
In the above we have used the following conventions and
notation. $\mathcal{K}_{ijk}$ and $\tilde{\mathcal{K}}_{abc}$ denote
the triple intersections on the manifold with SU(3) structure we use
in order to compactify the heterotic string and on its mirror
respectively. By $K_{cs}$ and $K_K$ we have denoted the parts in the
zeroth order K\"ahler potential, $K_0$, which depend on the complex
structure and K\"ahler moduli respectively. Finally, the symbols $\xi$,
$e_i$, $\epsilon_a$, $p_{ia}$, $\mu^a$, $q_i^a$, $\rho$ and $r_i$ are
the flux parameters. The Latin letters denote the fluxes due to the
$SU(3)$ structure while the Greek letters denote the usual $H$-fluxes.

The setup above is subjected to the following constraints
\begin{equation}
  \label{constr}
  \begin{aligned}
    p_{ia} q_j^a - p_{ja} q_i^a  - e_i r_j + e_j r_i =0 \; , \\    
    \xi r_i - \epsilon_a q_i^a  + \mu_a p_{ia} - \rho e_i = 0 \; , 
  \end{aligned}
\end{equation}
where the first of the constraints comes from the consistency
conditions on the manifold with $SU(3)$ structure while the second
comes from the Bianchi identity $dH = 0 $ which has to be satisfied in
the standard embedding case. The model above was derived in
\cite{GLM2} in a supergravity approximation and in order to insure its
validity one has to make sure that the moduli stay in the large volume
and large complex structure regime. Mathematically this translates
into the conditions
\begin{equation}
  \label{largeVcs}
  T^i + \bar T^i \gg 1 \; ; \qquad Z^a + \bar Z^a \gg 1 \; .
\end{equation}
On the other hand, the action for the charged fields is obtained
perturbative and this means it is valid only for matter fields which
are small fluctuations around the zero value
\begin{equation}
  \label{smallCD}
  |C^i| \ll 1 \; ; \qquad |D^a|  \ll 1 \; .
\end{equation}

Clearly, in general, the setup above is quite complicated as it
contains a large number of fields and parameters, and therefore it is
tedious to analyse. In order to make any progress we shall consider
that pairs of fields get large masses (due to the fluxes) and we can
integrate them out. We shall make this more precise in the next
section where we shall argue that the masses for (some of) the gauge
fields and those for the moduli are related and we can not integrate
out only moduli or matter fields.


\section{Masses for moduli and matter fields}
\label{masses}

In this section we study the masses for the moduli and matter fields
in the model presented previously. The motivation behind is that if we
find that certain fields acquire large masses, we can try to implement
a two step procedure in finding the vacuum, where in the first
instance we integrate out the heavy fields and then, in the second step we
find the vacua of the remaining theory. Therefore, if we have fluxes
which are generic enough, we can hope that the effective model we have
to analyse is sufficiently simple so that we can analyse it
analytically. 

Lt us assume that in the first step we are able to find a supersymmetric
solution which preserves the full $E_6$ gauge group. 
It is only a matter of algebra to compute the mass matrix. Starting
from the supergravity expression of the scalar potential
\begin{equation}
  \label{Vsugra}
  V = e^K \left( D_i W \overline{D_j W} g^{\bar \jmath i} - 3 |W|^2\right)
\end{equation}
one computes the second derivative of the potential as
\begin{eqnarray}
  \label{d2V}
  \partial_i \partial_j V & = & - e^K \partial_i \partial_j W \bar W -
  e^K \partial_i \partial_j K |W|^2 + e^K \partial_i K \partial_j K
  |W|^2 \; ; \nonumber \\
  \partial_i \partial_{\bar \jmath} V & = & -2 e^K g_{i \bar \jmath} |W|^2
  + e^K g^{\bar k l} \partial_{\bar \jmath} D_{\bar k} \bar
  W \partial_i D_l W \; ,
\end{eqnarray}
Where we have made use of the the fact that in a supersymmetric vacuum
\begin{equation}
  \label{susyv}
  D_i W = 0 \; .
\end{equation}
The indices $i, j, \ldots$ label all the chiral fields present
(including the charged matter fields) and should not be confused with
the indices $i, j, \ldots = 1 , \ldots h^{1,1}$ which were used in the
previous section  to label the K\"ahler moduli and the charged fields
in $\af$. In fact, in the following we shall come back to the notation
in the previous section where the indices $i, j , \ldots$ label the
K\"ahler moduli. 

The condition that the gauge group is
not broken ensures that the mass matrix splits into one for the
matter fields and one for the moduli. Moreover, in the matter field
mass matrix, only the gauge invariant terms survive. 
Splitting the indices into $T,Z$ for the moduli (K\"ahler and complex
structure respectively) and $C,D$ for the matter fields (in $\af$ and
$\f$ respectively) and suppressing extra indices on these fields in
order to avoid clutter, we find that the only non-vanishing terms of
the matter field mass matrix are
$\partial_C \partial_D V$, $\partial_C \partial_{\bar C} V$ and
$\partial_D \partial_{\bar D} V$. 
Computing these terms at the first order in $\alpha'$ we obtain
\begin{eqnarray}
  \partial_C \partial_D V & = & e^{-K} \left(\partial_C \partial_D
    W_1 \bar W_0 + \partial_C \partial_D K |W_0|^2 \right) \alpha' +
  \mathcal{O} ({\alpha'}^2) \; ; \nonumber \\
  \partial_C \partial_{\bar C} V & = & \left(-2 e^{K} g_{C \bar C}
  |W_0|^2 + e^{K_0} g^{D \bar D} \partial_{\bar C} D_{\bar D} \bar
  W_1 \partial_C D_D W_1\right) \alpha' + \mathcal{O} ({\alpha'}^2)  \; ; \\
  \partial_D \partial_{\bar D} V & = & \left( -2 e^{K} g_{D \bar D}
    |W_0|^2 + e^{K_0} g^{ \bar C C} \partial_{\bar D} D_{\bar C} \bar
    W_1 \partial_D D_C W_1 \right) \alpha' + \mathcal{O}
  ({\alpha'}^2)\; . \nonumber
\end{eqnarray}
Similar formulae can be written for the moduli fields
\begin{eqnarray}
  \partial_T \partial_T V & = & - e^K \partial_T^2 W \bar W -
  e^K \partial_T^2 K |W|^2 + e^K \partial_T K \partial_TK |W|^2 \; ; \nn \\
  \partial_T \partial_Z V & = & - e^{K} \partial_T \partial_Z  W \bar
  W - e^K \partial_T \partial_Z K |W|^2 + e^K \partial_T K \partial_Z
  K |W|^2 \; ; \nn \\
  \partial_Z \partial_Z V & = & - e^K \partial_Z^2 W \bar W -
  e^K \partial_Z^2 K |W|^2 + e^K \partial_Z K \partial_Z K |W|^2 \; ; \\
  \partial_T \partial_{\bar T} V & = & -2 e^K g_{T \bar T} |W|^2 + e^K
  g^{\bar T' T'} \partial_{\bar T} D_{\bar T'} \bar W \partial_T
  D_{T'} W + g^{\bar Z Z}\partial_{\bar T} D_{\bar Z} W \partial_T D_Z
  W \nn \; ; \\
  \partial_T \partial_{\bar Z} & = & e^K g^{\bar T' T'} \partial_{\bar
    Z} D_{\bar T'} \bar W \partial_T D_{T'} W + e^K g^{\bar Z'
    Z'} \partial_{\bar Z} D_{\bar Z'} \bar W \partial_T D_{Z'} W \nn
  \; ; \\
  \partial_Z \partial_{\bar Z} & = & -2 e^K g_{Z \bar Z} |W|^2 + e^K
  g^{\bar T T} \partial_{\bar Z} D_{\bar T} \bar W \partial_Z
  D_{T} W + g^{\bar Z' Z'}\partial_{\bar Z} D_{\bar Z'} W \partial_Z D_{Z'}
  W \; ; \nn
\end{eqnarray}

Note that the masses for the matter fields come at order $\alpha'$
while the masses for the moduli are only at zeroth order as usually
the $\alpha'$ corrections to the moduli are given by terms including
matter fields, which are taken to be zero in the background. 

So far the formulae written above are fairly general and do not make
use of the specific form of the superpotential or the K\"ahler
potential, apart from the fact that the K\"ahler potential is, at the
zeroth order in $\alpha'$, a sum of the K\"ahler potentials for the
K\"ahler moduli and for the complex structure moduli. In the following
we shall also make use of some specific properties 
of these quantities defined in \eqref{K0}, \eqref{K1}, \eqref{W0} and
\eqref{W1}. First note that $\partial_T \partial_T W = 0$ as the flux
superpotential comes linear in $T$. Then, for $E_6$ preserving
solutions $\partial_T \partial_Z K = 0$ as such terms appear
multiplied by matter fields which have to vanish in the
background. Finally, it is easy to see that
\begin{eqnarray}
  \label{Wcd}
  \partial_C \partial_D W_1
  \raisebox{-.25cm}{\rule{.3pt}{15pt}}_{\;C=D=0} & = &
  2 \partial_T \partial_Z W_0 \; ;  \nn \\
  \partial_C \partial_D K_1
  \raisebox{-.25cm}{\rule{.3pt}{15pt}}_{\;C=D=0} & = &  -2 \partial_T
  K_0 \partial_Z K_0 \; ; \\
  \partial_C \partial_{\bar C} K_1 
  \raisebox{-.25cm}{\rule{.3pt}{15pt}}_{\;C=D=0} & = & 4 e^{(K_{cs} -
    K_K)/3} g_{T \bar T} \; ; \nn \\
  \partial_D \partial_{\bar D} K_1 
  \raisebox{-.25cm}{\rule{.3pt}{15pt}}_{\;C=D=0} & = & e^{(K_K -
    K_{cs})/3} g_{Z \bar Z} \; , \nn 
\end{eqnarray}
which allow us to write all the masses only in terms of derivatives
of the zeroth order quantities $K_0$ and $W_0$.

There is one more thing we should note related to the formulae for the
mass matrix elements. The superpotential appears quadratically in each
term with or without derivatives. Since first order derivatives of the
superpotential can be replaced from \eqref{susyv} there will only be
terms involving two derivatives of the superpotential or plain
superpotential terms without any derivative. Moreover it is easy to
see that each plain superpotential factor comes together with a
derivative of the K\"ahler potential with respect to the moduli,
$K_X$, which, in the supergravity limit we are using, behaves like 
\begin{equation}
  K_X \sim \frac1{X + \bar X} \ll 1 \; .
\end{equation}
Therefore, if we are looking for a regime where the superpotential is
small, the leading terms in the mass matrix elements are the ones
which are quadratic in the second derivative of the
superpotential. Writing now for clarity the indices which label the
various fields -- but still ignoring the $E_6$ indices which are all
contracted as the results should be gauge invariant -- the relevant
elements of the mass matrix take the form
\begin{eqnarray}
  \label{mmw0}
  e^{-K} \partial_{C^i} \partial_{\bar C^{\bar \jmath}} V & = &  4 \alpha'
  e^{(K_{\mathrm{cs}} - K_K)/3} g^{\bar b a} \partial_{\bar
    \jmath} \partial_{\bar b} \bar W \partial_i \partial_a W +
  \mathcal{O}(W) \; ; \nn \\ 
  e^{-K} \partial_{D^a} \partial_{\bar D^{\bar b}} V & = &  \alpha'
  e^{(K_K -K_{\mathrm{cs}})/3} g^{\bar \jmath i} \partial_{\bar
    b} \partial_{\bar \jmath} \bar W \partial_a \partial_i W+
  \mathcal{O}(W) \; ; \nn \\
  e^{-K} \partial_{T^i} \partial_{\bar Z^{\bar a}} V & = & g^{\bar b
    a} \partial_{\bar a} \partial_{\bar b} \bar
  W \partial_i \partial_a W + \mathcal{O}(W) \; ; \\
  e^{-K} \partial_{T^i} \partial_{\bar T^{\bar \jmath}} V & = & g^{\bar b
    a} \partial_{\bar \jmath} \partial_{\bar b} \bar
  W \partial_i \partial_a W + \mathcal{O}(W) \; ; \nn \\
  e^{-K} \partial_{Z^a} \partial_{\bar Z^{\bar b}} V & = & g^{\bar d
    c} \partial_{\bar b} \partial_{\bar d} W \partial_a \partial_c W 
  + e^K g^{\bar \jmath i} \partial_{\bar b} \partial_{\bar \jmath}
  \bar W \partial_a \partial_i W  + \mathcal{O}(W) \; , \nn 
\end{eqnarray}
where the rest of the elements are order $\mathcal{O}(W)$ or higher.
It is now clear that the masses for the matter fields and those for the
moduli are related. In particular it can be seen that the masses are
mainly controlled by the mixed derivatives of the superpotential
$\partial_T \partial_Z W$. If this has zero eigenvalues, meaning that
the charged fields are massless, also the mass matrix for the
$T$-moduli will have zero eigenvalues. Therefore, massless matter
fields are only possible if some of the moduli remain flat directions.

One can explicitly check for a toy model which
contains only one $T$ and one $Z$ modulus that this is indeed the
case. For certain values of the flux parameters which give rise to 
a small $W$, \cite{dCGLM}, the masses for the matter fields turn out
to be suppressed only by the fact that they come in a higher order in
the $\alpha'$ expansion. Given that
in the setup we consider the $\alpha'$ scale is not much below the
Planck scale, any field which is massive can be safely removed form
the spectrum.

It should be clear that the above pattern can be used only for pairs
of fields $(T,Z)$ and $(C,D)$. The fields which remain ``outside''
such pairs are bound to have lower masses, of order $|W|$. This should
be clear for the unpaired matter fields, because by gauge invariance
and holomorphy of the superpotential the only possible second
derivatives of the superpotential involve both fields in $\f$ and
fields in $\af$. For the moduli fields this may not be totally correct
because of the existence of second derivatives of the superpotential
of the form $\partial_Z \partial_Z W$ which may give large masses to
the unpaired $Z$-fields. Nonetheless, the $T$-moduli are light.

Before we go further let us summarise the results of this section. We
found that in general the masses for the moduli and matter fields are
related and even though the latter may be lower than the former they
are still large and, in a low energy approximation, one should
truncate both the moduli and the corresponding matter fields. This
means that we can effectively remove from the spectrum $h =
\mathrm{min}(h^{1,1}, h^{2,1})$ pairs of moduli $(T, Z)$ and the
corresponding pairs of matter fields $(C, D)$. This will be our
strategy in the following in order to obtain simpler models which can
be more easily analysed.

\section{Effective low energy models}
\label{effective}

In the previous section we studied the masses for the matter fields
and for the moduli in a supersymmetric, $E_6$ preserving vacuum. Even
though in the general case it is hard to perform explicit
calculations, we argued that the masses for the matter fields and for
the moduli are related. We now apply these findings and truncate from
the spectrum $h= min (h^{1,1} h^{2,1})$ pairs of moduli, $T,Z$ and the
corresponding matter fields $C,D$. Note that we are not doing a proper
integration out of the heavy fields, but rather freeze them at the
value they have at the supersymmetric solutions analysed in the
previous section. The conditions under which such a procedure is
consistent were analysed in \cite{GS1,GS2}. As for the
case at hand we do not know the solution explicitly these conditions
are hard to analyse. We shall nevertheless assume that freezing the
heavy fields at the supersymmetric value is consistent.

In the following, w+e shall distinguish two cases
\begin{enumerate}
\item $h^{1,1} > h^{2,1}$,
\item $h^{2,1} > h^{1,1}$.
\end{enumerate}
In the first case we assume that $h^{2,1}$ pairs of moduli $(T,Z)$ get
large masses and can be removed from the spectrum. The same will
happen with $h^{2,1}$ pairs of matter fields $(C,D)$ and we are
effectively left with a model containing only K\"ahler moduli and the
corresponding charged fields $C$ in the $\overline{\mathbf{27}}$ of
$E_6$. In the second case the situation will be opposite and we will
be left with an effective model with only complex structure moduli and
the corresponding matter fields, $D$, in $\mathbf{27}$ of $E_6$.

Before we discuss each case let us comment on the constraints one
would have to satisfy in these effective models. In general we need to 
choose the flux parameters such that the constraints \eqref{constr}
are satisfied. However, this would be relevant if we solved the
equations of motion in full generality. What we want to do instead is
to see the effect of the constraints \eqref{constr} on the effective
models we want to analyse.

The important point to
note is that these constraints are non-trivial only if both electric
and magnetic type fluxes are present. For example if the
superpotential \eqref{W0} contained only up to linear terms in the
complex structure moduli, the constraints
\eqref{constr} would be trivial. Also, if the superpotential did not
depend on the K\"ahler, or on the complex structure moduli, the
constraints \eqref{constr} would be again identically
satisfied. Therefore it is likely that these constraints do not have a
descendant in the effective models we have listed above as they only
contain one type of moduli fields (either K\"ahler or complex
structure moduli). 

We can nevertheless expect that the constraints
\eqref{constr} play a role in the process of integrating out the
massive fields. One thing which may happen is that the constraints
enforce that the masses take particular values which may not be
consistent with the integrating out procedure we want to
implement. In particular it would be disturbing if the constraints
imply that the masses for certain fields we wanted to integrate out,
vanish. As we have seen in the previous section the masses for the
fields which pair up are controlled by the matrix
$\partial_T \partial_Z W$, which in such a case would have zero
eigenvalues. By looking at a system which contains only one pair of
moduli, $T,Z$, and one pair of matter fields $C,D$, this does not seem
to be the case so in the following we will assume that the constraints
do not play a role in the values of the masses for the paired up fields. 

One other obstruction may be that the constraints \eqref{constr} do
not allow values for the moduli within the region where the
approximations we use are valid. Some simple computer scan over the
parameter space shows that actually this is not the case. Moreover, in
section \ref{matterw} we shall see that there is enough freedom in the
parameters to fix the values of the moduli in the region of validity
for our approximations. As we shall learn in this section this can be
done at the expense of not having the value of the superpotential as a
free parameter anymore. In particular we shall learn that the
superpotential can not be made zero while keeping the moduli fields in
a region where the approximations we use still hold.

In conclusion, we assume that we can fix the moduli at values
consistent with all the approximations we use and that we can give
them large enough masses so that they are effectively removed from the
spectrum. Moreover, we assume that the effective models listed above
do not have to satisfy any constraint like \eqref{constr}.  The only
restriction will be on the value of the superpotential which will
actually appear as a constant in the effective superpotential obtained
after integrating out the massive fields. This term will be in general
large (at least of order one) and since we can not control it very
well we should avoid moduli stabilisation mechanisms which make use of
such a term.

\subsection{$h^{1,1} > h^{2,1}$}

In this case we will be effectively left over with $h^{1,1} - h^{2,1}$
K\"ahler moduli and the corresponding matter fields in $\af$. The
superpotential has the form 
\begin{equation}
  \label{WK}
  W = w_0 + eT + C^3 \; ,
\end{equation}
and it was analysed in \cite{AM}. In the case that $w_0 =0$ the system
was shown to have no proper solution with fixed moduli. In the case
$w_0 \ne 0 $ one can find a $E_6$ solution and the value of the
K\"ahler modulus will directly depend on the (complex) parameter
$w_0$. Unless one does a proper integration over the massive fields
which were removed form the spectrum, the constant $w_0$ is unknown,
and it is not clear what would be the values for this constant which
are compatible with all the constraints on the theory. Finally, the
value for the superpotential in such cases is typically large and
incompatible with the stabilisation of the dilaton via a gaugino
condensate in the hidden sector.

\subsection{$h^{2,1} > h^{1,1}$}

This case is similar to the previous one, with the difference that now
we are dealing with complex structure moduli rather than K\"ahler
moduli and accordingly, the matter fields transform in the $\f$
of $E_6$. Compared to the previous case, the superpotential for the
complex structure moduli is more complicated and allows for more
flexibility in fixing the moduli
\begin{equation}
  \label{Wcs}
  W= w_0 + \epsilon Z + \frac{\mu}{2} Z^2 + \frac{\rho}{3} Z^3 + D^3
  \; .
\end{equation}
As we said before, we are looking for solutions which have a small
value for the superpotential at the critical point so that in the end
we can use the gaugino condensate in order to stabilise the
dilaton. To gain a qualitative picture, note that such solutions are
given in a first approximation by solving the global supersymmetry
equations, $\partial_Z W =0$ (see \cite{dCGLM} for details). For the
case at hand, this equation can be easily solved and one can also
compute the value of the superpotential at this point. With enough
flexibility in tuning the flux parameters, one can achieve a small $Re
\; W$, but it can be easily seen that the imaginary part is
proportional to $ z^3$. The consistency of the model requires that $z
\gg 1$ and therefore it is impossible to achieve a small value for the
superpotential in this regime. In conclusion, even though the starting
superpotential \eqref{W0} looks complex enough so that one may naively
think that it is easier to tune the parameters such that a small value
for $W$ is achieved, in practice, we have argued that this does not
happen and we encounter the old problem that
we have to balance $W_{flux}$, which is of order one or greater due to
flux quantization, against the non-perturbative small effect of gaugino
condensation \cite{DIN2,DRSW,RW,DIN1}.

\section{Breaking $E_6$}
\label{matterw}

The strategy we followed so far was to integrate out the massive fields
and analyse the simpler models obtained in this way. We have seen in
the previous section that there may be solutions to these simple
models, but in general the value for the superpotential is large and
incompatible with the stabilisation of the dilaton via gaugino
condensation in the hidden sector. Small values for $W$ are also
needed in order to be sure that the truncation of the massive modes, we
have done in the first step, is indeed consistent. In this section we
shall investigate another possibility of obtaining small values for
the superpotential.

Since the fluxes are integers, the solutions for the moduli fields
will generically be rational numbers. Therefore the superpotential
will be a polynomial with integer coefficients of these rational
numbers. It is more or less clear that unless one chooses a big
hierarchy between the fluxes, arbitrarily small, but nonvanishing
values for the superpotential are not possible. However, from rational
and integer numbers it is much easier to construct a vanishing
quantity. Still, we need a non-vanishing superpotential in order to be
able to fix the dilaton via gaugino condensation in the hidden
sector. This can be achieved in principle by vevs of the matter
fields. If the values for the charged fields are small -- which is in
fact required by the consistency of the derivation of the effective
action in the presence of charged fields -- the superpotential will
naturally be a small quantity. We shall see in this section that
actually such solutions are not possible as they violate the
constraint on the fluxes coming from the Bianchi identity. 

Note that this idea is not entirely new, but it appeared before in the
literature in various forms \cite{DIN2,RW,GKLM}. In these
references it was suggested that the small quantity which should
balance the gaugino condensate comes from the Chern-Simons correction
to the field strength $H$ of the antisymmetric tensor field. Recall that
for the case at hand, the superpotential is generated from the formula
\cite{GLM1} 
\begin{equation}
  W = \int H \wedge \Omega \; ,
\end{equation}
where $\Omega$ is the $(3,0)$ form which one can define on a manifold
with $SU(3)$ structure. Moreover, the matter superpotential comes
entirely from the the Chern-Simons correction to $H$ which shows the
relation with the references above. The difference now is that for the
case at hand we have a very specific model where moduli stabilisation
can be discussed explicitly and we can check whether such a mechanism
works or not. 

\subsection{Preliminaries}

For the model described in section \ref{model}, suppose we are given a
supersymmetric solution in flat Minkowski space at zeroth order in
$\alpha'$. The purpose is to find a solution at order $\alpha'$ which
has non-vanishing superpotential which can be used later to fix the
dilaton via gaugino condensation in the hidden sector. At first order
in $\alpha'$ the equations for supersymmetric solutions read
\begin{equation}
  \label{eq:1}
  \partial_C W_1 = \partial_D W_1 =0 \; .
\end{equation}
In the above equations, the term proportional to the superpotential is
absent as $W_0$ vanishes by assumption while the term with $W_1$ is
higher order in $\alpha'$. A non-vanishing $W_1$ can only be obtained
if the charged fields have a nonzero vev which in turn implies that
the $E_6$ gauge group is broken. The equations above can be written
schematically as
\begin{equation}
  \label{C2D}
  j C^2 + D =0 \; , \qquad j D^2 + C = 0 \; ,
\end{equation}
where $j$ denotes the totally symmetric cubic invariant of $E_6$.
Let us make a few comments here. First of all, note that the presence
of the mass term $CD$ in the superpotential is crucial as, in its
absence, only $E_6$ preserving solutions can be found. Moreover, the
range of the gauge groups which are preserved by the solutions to the
above equations is limited and can not be any subgroup of $E_6$. For
example let us assume that we are looking for a solution which
preserves an $SO(10)$ gauge group. The $\mathbf{27}$ of $E_6$
decomposes under $SO(10)$ as
\begin{equation}
  \label{27}
  \mathbf{27} = \mathbf{1} \oplus \mathbf{10} \oplus \mathbf{16} \; ,
\end{equation}
and therefore this breaking would be triggered by a non-vanishing vev
of the singlet field in the above decomposition. It is easy to see
that in this case the first term in \eqref{C2D} does not contribute
because of the coefficient of the cubic coupling of the singlet field
vanishes, ie $j_{111}=0$. This means that the only solution will be the
trivial one which actually preserves the full $E_6$ group and not only
a $SO(10)$ subgroup.

Even if, as noticed above we do not have full flexibility in choosing
a solution to the above equations, reasonable solutions with $C, D \ll
1$ to these equations may be found depending on the combinatorial and
group theoretical factors. We do not insist any further on this aspect
as the purpose of this preliminary section was only to show that
apriori the line of reasoning we have chosen can indeed work. We still
need to check that the initial assumption of the existence of a
Minkowski ground state at the zeroth order in $\alpha'$. We shall
actually prove in the following that this is not possible in the setup
we consider.

\subsection{No-go theorem for Minkowski solutions}

Now let us look for Minkowski solutions in the low energy theories
obtained from heterotic string compactifications on manifolds with
$SU(3)$ structure with standard embedding. A computer scan of the flux
parameter space shows that at zeroth order in $\alpha'$, many
solutions with vanishing superpotential exist with moderate values for
the fluxes. However none of these solutions satisfy the constraint on
the fluxes imposed by the Bianchi identity. It is important to stress
that this constraint has to be imposed as we have already chosen, by
embedding the spin connection into the gauge group, that the right
hand side of the Bianchi identity
\begin{equation}
  \label{BI}
  dH = tr (F\wedge F - R \wedge R) \; ,
\end{equation}
identically vanishes. Therefore we have to make sure that for the
solutions we find, $H$ satisfies $dH=0$.

In order to be able to manage the calculations let us choose a setup
which has one K\"ahler and one complex structure modulus. Together
with these fields we have to consider the corresponding charged fields
which we have briefly analysed in the previous subsection. As we
mentioned there, the term $CD$ in the superpotential is crucial for
finding solutions, and this is why we need both matter fields in $\f$
and in the $\af$ of $E_6$. The zeroth order (in $\alpha'$)
superpotential for this model is given by
\begin{equation}
  \label{WTZ}
  W = i(\xi + i e T) + (\epsilon + ipT)Z + \tfrac{i}2 (\mu + iqT) Z^2 +
  \tfrac16 (\rho + irT) Z^3 \; ,
\end{equation}
where $T$ and $Z$ are the K\"ahler and respectively complex structure
moduli fields and $\xi$, $\epsilon$, $ \mu$, $\rho$, $e$, $p$, $q$,
$r$ are the flux parameters which satisfy the constraint \cite{GLM1}
\begin{equation}
  \label{con}
  \xi r - \epsilon q + \mu p - \rho e =0 \; .
\end{equation}
The flux parameters are in principle quantised in certain units, but
in the following we won't even need this additional requirement and we
shall only use the fact that they are real numbers (ie are not
complex).

Supersymmetric Minkowski solutions to this setup are given by the
equations
\begin{eqnarray}
  \label{WT}
  \partial_T W & = & -e + ipZ -\tfrac{q}2 Z^2 + \tfrac{i}6 r Z^3 = 0
  \; , \\
  \label{WZ}
  \partial_Z W & = &(\epsilon + ipT) + i (\mu + iqT) Z + \tfrac12 (\rho +
  irT) Z^2 = 0 \; , 
\end{eqnarray}
supplemented by the condition $W=0$.

The strategy is fairly clear: the equation \eqref{WT} is cubic and can
be solved for $Z$. Then one replaces this solution into \eqref{WZ} in
order to determine $T$. Finally, we have to impose the constraints
\eqref{con} and $W=0$ which will restrict the flux parameter
space. Apriori one should think that there are enough flux parameters
which we can tune such that we get the desired solution, but as we
shall see in the following, this intuition is not correct.

It is useful to remark the following trick which sometimes eases the
calculations. Let us denote
\begin{equation}
  \label{TZ}
  T= t + i \tau \; , \qquad Z = z + i \zeta \; ,
\end{equation}
where $t$ and $z$ are the true moduli fields, ie the moduli which
govern the actual size of the corresponding cycles, while $\tau$ and
$\zeta$ are their axionic superpartners. With this notation, it is a
matter of algebra to derive
\begin{equation}
  \label{eq:4}
  Re \left( t \partial_TW + z \partial_Z W - W \right) = z^2 (-qt +
  \tfrac13 \rho z - r t \zeta - \tfrac13 r z \tau)  \; .
\end{equation}
We are looking for a solution for which each of the terms on the left
hand side vanishes and since the approximations we are using require
that $z \gg 1$ (which in particular means that $z\ne0$) we find that
\begin{equation}
  \label{ReW}
  -qt + \tfrac13 \rho z - r t \zeta - \tfrac13 r z \tau =0 \; ,
\end{equation}
which is a condition much easier to analyse than the generic $Re \; W
=0$.

In the following we shall distinguish two cases which are
intrinsically different: $r\ne 0$ and $r=0$.

\subsubsection{$r \ne 0$  case}

Let us start with some simple observation. Let us shift the field $Z$
by some purely imaginary constant 
\begin{equation}
  \label{eq:5}
  Z = Z' + ia \; , \quad a \in  \mathbf{R} \; .
\end{equation}
Defining new flux parameters by
\begin{equation}
  \label{eq:3}
  \left(
    \begin{array}{c}
      e' \\
      p' \\
      q' \\
      r'
    \end{array} \right) = M 
  \left(
    \begin{array}{c}
      e \\
      p \\
      q \\
      r
    \end{array} \right) \quad  \mathrm{and} \quad
  \left(
    \begin{array}{c}
      \xi' \\
      \epsilon' \\
      \mu' \\
      \rho'
    \end{array} \right) = M 
  \left(
    \begin{array}{c}
      \xi \\
      \epsilon \\
      \mu \\
      \rho
    \end{array} \right) \; ,
\end{equation}
where the matrix $M$ is defined as
\begin{equation}
  \label{eq:6}
  M = \left(
    \begin{array}{c c c c}
      1 & a & -\tfrac12 a^2 & - \tfrac16 a^3 \\
      0 & 1 & -a & - \tfrac12 a^2 \\
      0 & 0 & 1 & a \\
      0 & 0 & 0 & 1 
    \end{array} \right) \; ,
\end{equation}
the superpotential \eqref{WTZ} has precisely the same form, but with
the primed quantities. In matrix notation the constraint reads
\begin{equation}
  \label{conmat}
  (e,p,q,r) L
  \left(
    \begin{array}{c}
      \xi \\
      \epsilon \\
      \mu \\
      \rho
    \end{array} \right) = 0 \; ,
\end{equation}
where $L$ is the simplectic form
\begin{equation}
  \label{eq:8}
  L =  \left(
    \begin{array}{c c c c}
      0 & 0 & 0 & 1 \\
      0 & 0 & -1 & 0 \\
      0 & 1 & 0 & 0 \\
      -1 & 0 & 0 & 0
    \end{array} \right) \; .
\end{equation}
It is easy to verify that $M^T L M = L$ which means that the
constraint also has the same form in the primed parameters.

The next thing to notice is that the equation which determines $Z$,
\eqref{WT}, always has (at least) one purely imaginary solution which
we denote by $Z_0$. To
see this note that making the change $Z \to iZ$ changes the equation
in a cubic equation with real coefficients which always admits a real
solution. 

Now let us combine the above remarks and make a shift in $Z$ by the
purely imaginary solution of \eqref{WT}.\footnote{Note that we are
  interested in setups which give consistent values for the moduli, ie
  $Re Z \ne 0$. This means that equation \eqref{WT} should have
  precisely one purely imaginary solution which makes the above shift
  unambiguous.}
In the notation above it
means that $a = -i Z_0$. The advantage is that in the
new (primed) variables the flux parameter $e' = e -i p Z_0 + \tfrac12
q Z_0^2 - i \tfrac16 r Z_0^3 \equiv 0$.

By the above tricks we have ended with precisely the same theory we
started from, but with one vanishing flux parameter. Note that the
shift in $Z$ is not of any physical significance, as the imaginary part
of $Z$ was an axion before turning on fluxes, and its vev does not appear
in the calculation of any physical quantities. If, on the other hand,
the same trick had required a shift in the real part of $Z$ instead, 
we would have had to be more careful as the value of this field is
important in the consistency of the setup.

Finally, we should stress that the key ingredient here is the fact that
the fluxes are real parameters. This is unlike type IIB theory where
the flux superpotential contains complexifications of the flux
parameters and so the above argument can not hold. 

With this comments we shall start analysing the solutions of the
system defined by the equations \eqref{WT} and \eqref{WZ} subject to
the constraints $Re \; W =0$ (which is equivalent to \eqref{ReW}), $Im
\; W =0$ and \eqref{con}. We shall consider that $e=0$ which we can
do, as explained above, without loss of generality.

Equation \eqref{WT} can now be solved easily. The solution $Z=0$ is
not physical as we need for consistency that $z \equiv Re \; Z \gg
1$. We find that
\begin{equation}
  \label{eq:7}
  \zeta = - \frac{3q}{2r} \quad \mathrm{and} \quad z^2 = -
  \frac{9q^2}{4 r^2} - 6 \frac{p}{r} \; .
\end{equation}
Note that this solution is valid only if the expression defining $z^2$
is positive. Therefore we necessarily have $p\ne 0$.

To solve for $T$ is is easier to use the constraint \eqref{ReW}. From
this we can express $\tau$ as a function of $t$, $z$ and $\zeta$ and
replace the result into $Im \; \partial_Z W$. After a little algebra,
making use of the solution for $Z$ we find
\begin{equation}
  \label{toz}
  \frac{t}{z} = \frac{\mu  - \rho q}{2rp} \; .
\end{equation}

At this stage we have found a solution for $T$ and $Z$ and we have to
impose the constraints coming from $Re \; \partial_Z W=0$, $Im \; W=0$
and \eqref{con}. The strategy will be to express $\epsilon$ and $\xi$
from the first and last equations and replace these results together
with the solution for $T$ and $Z$ in $Im \; W$. After straightforward
algebraic calculations one finds
\begin{equation}
  \label{ImW}
  Im \; W = \tfrac43 ptz \; .
\end{equation}
The consistency of the setup requires that $t$ and $z$ be
non-zero. Moreover, we have explained above that a non-vanishing
solution for $z$ is possible only for $p \ne 0$ and therefore the
right hand side of the above equation can not vanish. This ends our
proof that for $r\ne 0$ the setup we have considered has no
supersymmetric Minkowski solutions which satisfy the constraint
\eqref{con}.

\subsubsection{$r=0$ case}

As we have explained at the beginning of this section, the
computations above apply only to the case $r \ne 0$ and for vanishing
$r$ we have to redo all the calculations. First of all note that if
$r=0$, the equation \eqref{WT} no longer has a purely imaginary
solution by which we can shift $Z$. Therefore our argument that we can
shift $Z$ in order to make $e=0$ no longer works. In fact, in order to
have a non-vanishing solution for $z$ we need that $p^2 + 2 qe <0$
which actually requires that both $q$ and $e$ are non-zero. Then the
solution to \eqref{WT} reads
\begin{equation}
  \label{eq:9}
  \zeta = \frac{p}{q} \quad \mathrm{and} \quad z^2 = - \frac{p^2}{q^2}
  - 2 \frac{q}{e} \; .
\end{equation}
The equation for $T$, \eqref{WZ}, can easily be solved to yield
\begin{equation}
  \label{eq:10}
  \tau = \frac{\rho p + \mu q}{q^2} \quad \mathrm{and} \quad t =
  \frac{1}{qz} \left( \epsilon - \frac{\mu p}{q} - \frac{\rho e}{q} -
    \frac{p^2 \rho}{q^2} \right) \; .
\end{equation}
The first two terms in the bracket can be replaced from the constraint
\eqref{con} and one easily finds
\begin{equation}
  \label{eq:11}
  \frac{t}{z} = \frac{\rho}{q} \; .
\end{equation}

Looking at the constraint coming from $Re \; W=0$, \eqref{ReW}, for
the case $r=0$ we see that the two equations are not compatible with\
each other, and so, also for the case $r=0$ there is no Minkowski
solution which satisfies \eqref{con}. In the above it seems that we
have not even made use of the constraint coming from $Im \; W =0
$. This is indeed true, but this constraint can always be satisfied by
choosing the flux parameter $\xi$ accordingly. Moreover, this flux
parameter does not modify the constraint \eqref{con} because
$r=0$. The only problem may come from the fact that fluxes are
quantised, but by running a simple computer code to scan over the flux
parameter space it can be seen that there exist several solutions which
have $\xi$ integer.

This ends our proof that in heterotic models compactified on manifolds
with $SU(3)$ structure with standard embedding there are no
supersymmetric Minkowski solutions with all moduli fixed. Therefore
the trial to find solutions which have small $W$ which is generated
entirely from the matter sector fails. 

We should close this section by one remark. The proof we have
presented above is entirely at the zeroth order in $\alpha'$ and is
independent of the presence of matter fields. Therefore one would have
expected that such an analysis had appeared before in the
literature. Indeed, this problem was studied before in
\cite{dCGLM}. There solutions with small $W$ were found by perturbing
global supersymmetric solutions (which are the same as Minkowski
solutions). The reason such solutions were found was precisely because
the Bianchi identity was not imposed. The consistency there was argued
via the possibility of adding NS5 branes in order to satisfy the
Bianchi identity. For the case at hand this is not a possible solution
anymore because in order to deal with matter fields we have to specify
our solution to the Bianchi identity in the first place (ie by
standard embedding in this case) and the matter field spectrum only
follows after that. We have chosen the standard embedding because we
know how to treat the matter fields and this means that the Bianchi
identity can not be modified any further by the inclusion of NS5 branes.

\section{Conclusions and outlook}
\label{conclusions}

In this paper we have analysed the issue of moduli stabilisation in
models derived from heterotic strings compactified on manifolds with
SU(3) structure at first order in $\alpha'$. The setup used was the
one derived in \cite{GLM2} where the solution to the Bianchi identity
was obtained via the standard embedding. However, this setup suffers
from various problems, some of which can be traced back to the original
discussions about moduli stabilisation in heterotic
compactifications \cite{DIN2,DRSW,RW,DIN1}. One of the problems which has
been observed in this paper is that the moduli and matter field
sectors are not independent, but the masses in these sectors are the
same in many cases of interest. This can be directly related to the
fact that the moduli and matter fields arise from expansions in the
same set of forms on the internal manifold in the standard embedding
case. Therefore, giving masses to some of the moduli by making some of
the expansion forms non-harmonic, immediately implies that the
corresponding charged fields are also massive removing them from the
spectrum. Thus the first question to ask is how one can be left with
massless matter fields in the 4d theory and at the same time give
masses to all the moduli. One answer is that gauge symmetry prevents
the matter fields which transform in complex representations of the
gauge group to acquire masses unless pair anti-matter fields are
present. Such cases were discussed in section \ref{effective} and the main
conclusion is that the remaining moduli superpotential is too simple
in order to have a satisfactory picture of moduli stabilisation. In
particular, in the case where there are more K\"ahler than complex
structure moduli, one can argue along the lines of \cite{AM} that
there is no solutions with all moduli fixed, while in the other case,
where there are more complex structure than K\"ahler moduli, the
superpotential is too large in order to have the stabilisation of the
dialton via gaugino condensate in the hidden sector. 

Finally we have analysed one possibility of obtaining a small
superpotential at the minimum by finding a Minkowski flux solution (ie
vanishing superpotential at order $\alpha'{}^0$) and generating a
small superpotential by giving small vevs to the matter fields and
thus breaking the original $E_6$ gauge group. Even if apriori it is
not clear why such a setup can not exist, we have been able to prove
analytically that Minkowski solutions to the flux superpotential at
zeroth order in $\alpha'$ which obey the standard embedding Bianchi
identity do not exist. A similar analysis of moduli stabilisation was
performed also in \cite{dCGLM} where a solution was found precisely by
relaxing the Bianchi identity constraint which can be achieved by
introducing NS5 branes. In order to deal with the matter fields,
however, one has to specify an explicit solution to the Bianchi
identity, which can no longer be modified at will when dealing with
matter fields as the way the matter fields are defined intrinsically
depends on the solution to the Bianchi identity. Therefore, if we
want to modify the constraint which comes from the Bianchi identity by
adding branes, we would have to modify the matter sector accordingly
and so, the model from which we start would be different from
the one we have considered in this paper. There exist another logical
possibility of obtaining a consistent solution in the framework of
section \ref{matterw}. There we have considered non-trivial vevs for
the matter fields and wanted to use them in order to obtain a small
superpotential. The vacuum expectation values for the matter fields
modify the anomaly cancellation condition. 
Therefore the constraint \eqref{con} would strictly speaking have some
non-vanishing correction on the right hand side. This correction is
however small because we know how to treat only the matter fields
which are small fluctuations. This correction would show up as a
correction to \eqref{ImW}, but this can not compensate for the large
contribution on the RHS of \eqref{ImW} because, as explained before,
$p$ is quantised and therefore can not be made arbitrarily small,
while for consistency we need that $t, z \gg 1$.

It looks like all the problems encountered in this communication point
towards the fact that in heterotic string compactifications with standard
embedding moduli can not be stabilised in a satisfactory way. One
possible direction is to study heterotic string compactifications with
fluxes outside the standard embedding scenario. Another possibility is
to consider generalisations which include non-geometric backgrounds,
but both these directions have been unexplored so far.

\textbf{Acknowledgments} 
This work was supported in part by the National University Research
Council (CNCSIS) under the reintegration grant 3/3.11.2008 and program
"Idei", contract number 464/15.01.2009.

\end{document}